\documentclass[12pt,single]{article} 
\usepackage{anysize}

\usepackage[dvips]{graphicx} 
\usepackage{amsmath}
\usepackage{amssymb} 
\usepackage{amsfonts}
\usepackage{dcolumn}
\usepackage[hang,small,bf]{caption}

\marginsize{2.5cm}{2.5cm}{2.1cm}{2.5cm}  
\begin{document}

\begin{center}
\section*{Lamb   Shift of Muonic Deuterium  and Hydrogen }
\end{center}    

\subsection*{E. Borie}

\subsubsection*{Forschungszentrum Karlsruhe,      \\                     
Institut f\"ur Hochleistungsimpuls and Mikrowellentechnik (IHM), \\ 
Hermann-von-Helmholtzplatz~1,\\ 76344 Eggenstein-Leopoldshafen, Germany}


\vspace{1.2cm}

\subsubsection*{Abstract}
\vspace{-0.2cm}
My previous calculations of the Lamb shift in muonic hydrogen are  
reviewed and compared with other work.  In addition, numerical results 
for muonic deuterium are presented.



\subsubsection* { Introduction }  
\vspace{-0.2cm}
The energy levels of muonic atoms are very sensitive to effects of
quantum electrodynamics (QED), nuclear structure, and recoil, since the
muon is about 206 times heavier than the electron \cite{RMP}.  A number 
of theoretical analyses of the Lamb shift (the 2p-2s transition) in
light muonic atoms have been published  
\cite{eides,Borie05,Pachuki1,Pachuki2,Borie75,Borie78,hyperfine,digiacomo,BorieHe3},  most recently in view
of a proposed measurement of the Lamb shift im muonic hydrogen
\cite{experiment}.
The present paper repeats the independent recalculation of some of the
most important effects \cite{Borie05} and extends the numerical
calculations to the case of muonic deuterium, including effects that
were not considered previously \cite{carboni}.
Muonic deuterium is in many ways similar to muonic hydrogen, but there
are some differences.  In addition to the different mass 
the deuteron has spin 1 and both magnetic and quadrupole moments.  


In the numerical calculations the fundamental constants from  CODATA
 2002  (\cite{codat02}): 
$\alpha^{-1}$, $\hbar c$, $m_{\mu}$, $m_e$, $m_u$\,=\,137.0359991,
 197.32697\,MeV$\cdot$fm, 
 105.658369\,MeV, \\
 0.5109989\,MeV,  931.5050\,MeV, respectively     

Also, the following properties of the proton and deuteron were used:
$m_p$\,=\,938.272\,MeV/c$^2$, $R_p$\,=\,0.875\,$\pm$\,0.007\,fm 
(other recent values are discussed below) and  
$\mu_p\,=\,2.79285\,\mu_N$.  Also, 
 $m_d$\,=\,1875.613\,MeV/c$^2$, $R_d$\,=\,2.139\,$\pm$\,0.003\,fm  and 
$\mu_d\,=\,0.85744\,\mu_N\,=\,0.307012\,\mu_p$. (\cite{codat02})  
The deuteron has spin 1 and thus has both magnetic and
quadrupole moments.  The quadrupole moment of the deuteron is taken 
to be Q\,=\,0.2860(15)\,fm$^2$~\cite{friar02,reid-v,bishop}. \\




\subsubsection*{Vacuum Polarization }
\vspace{-0.2cm}
The most important QED effect for muonic atoms is the virtual production
and annihilation of a single $e^+ e^-$ pair 
It has as a consequence an effective interaction of 
order $\alpha Z \alpha$ which
is usually called the Uehling potential (\cite{Uehling,serber}.  This
interaction describes the most important modification of Coulomb's law.
Numerically it is so important that it should not be treated using
perturbation theory;  instead the Uehling potential should be added to
the nuclear electrostatic potential before solving the Dirac equation.
However, a perturbative treatment is also useful in the case of very
light atoms, such as hydrogen.  However, 
unlike some other authors, we prefer to  use relativistic (Dirac) wave
functions to describe  the muonic orbit.  Since these contributions have
been extensively discussed in the literature
\cite{RMP,eides,Borie05,Pachuki1} (among others), there is no need to go
into detail here.  The results, calculated  as the
expectation value of the Uehling potential using point-Coulomb Dirac
wave functions with reduced mass are, for muonic deuterium:
\begin{center} 
    \begin{tabular}{|l|cc|cc|}
   \hline
 &                  point nucleus  &        &    $R_d$=2.139\,fm & \\
 \hline
 & $2p_{1/2}-2s_{1/2}$  & $2p_{3/2}-2s_{1/2}$  & $2p_{1/2}-2s_{1/2}$  &
 $2p_{3/2}-2s_{1/2} $           \\
Uehling          &  227.6577~  & 227.6635~ &   227.5985~ &  227.6043~  \\
Kaellen-Sabry   &     1.66622  &   1.66626  &    1.66577  &  1.66582   \\
   \hline
\end{tabular}        
\end{center}   
The effect of finite proton size calculated here can be  parametrized as
-0.0129$\langle r^2 \rangle$. 
However higher iterations can change these results.  Up to now, these 
have not been calculated well for muonic deterium, as far as I know.

Corresponding numbers for muonic hydrogen, calculated as the
expectation value of the Uehling potential using point-Coulomb Dirac
wave functions with reduced mass are:
\begin{center} 
    \begin{tabular}{|l|cc|cc|}
   \hline
 &                  point nucleus  &        &    $R_p$=0.875fm  &  \\
 \hline
 & $2p_{1/2}-2s_{1/2}$  & $2p_{3/2}-2s_{1/2}$  & $2p_{1/2}-2s_{1/2}$  &
 $2p_{3/2}-2s_{1/2} $           \\
Uehling          &  205.0282~  & 205.0332~ &   205.0199~ &  205.0250~  \\
Kaellen-Sabry   &    1.50814  &   1.50818  &    1.50807  &  1.50811   \\
   \hline
\end{tabular}        
\end{center}   
The effect of finite proton size calculated here can be  parametrized as
-0.0109$\langle r^2 \rangle$. 
However higher iterations can change these results.  
The contribution due to two and three iterations have been calculated by
\cite{Pachuki1} and \cite{kinoshita}, respectively, giving a total of
0.151\,meV.  An additional higher iteration including finite size and
vacuum polarization is given in ref.\,\cite{Pachuki1} (equations(66) and
(67))  and ref.\,\cite{eides} (equations(264) and (268)).  These amount
to  -0.0164$\langle r^2 \rangle$.    
The best way to calculate this would be an accurate
numerical solution of the Dirac equation in the combined Coulomb-plus
Uehling potential.    

The mixed muon-electron vacuum polarization correction
(\cite{HelvPA,eides}) is  0.00007\,meV for hydrogen and 
 0.00008\,meV for deuterium. 
  
The Wichmann-Kroll  contribution was calculated using the
parametrization for the potential given in \cite{RMP}.  The result
obtained for hydrogen is -0.00103\,meV, consistent with that given in
\cite{eides}.  For deuterium, the contribution is  -0.00111\,meV. 
 
The equivalent potential for the virtual Delbr\"uck effect was
recomputed from the Fourier transform given in  \cite{Borie76} and
\cite{RMP}.  The resulting potential was checked by reproducing
previously calculated results for the 2s-2p transition in muonic Helium,
and the 3d-2p transitions in muonic Mg and Si.  The result for hydrogen
is +(0.00135 $\pm$ 0.00015)\,meV, and  for deuterium it is 
 +(0.00147 $\pm$ 0.00016)\,meV.   As in the case of muonic helium, this
contribution very nearly cancels the Wichmann-Kroll contribution.  
The contribution  corresponding to three photons to 
the muon and one to the proton  should be analogous to the light by light 
contribution to the muon anomalous moment; to my knowledge, the
corresponding contribution to the muon form factor has never been
calculated.   It will be comparable to the other light by light 
contributions.  This graph was included in contributions to the muon's
 anomalous magnetic moment; the contribution to the muon form factor is
one of the most significant unknown corrections. 

The sixth order vacuum polarization corrections to the Lamb shift in
muonic hydrogen have been calculated by Kinoshita and Nio
\cite{kinoshita}.  Their result for the 2p-2s transition (in hydrogen) is 
\begin{displaymath}
\Delta E^{(6)} ~=~ 0.120045 \cdot (\alpha Z)^2 \cdot m_r
\left(\frac{\alpha}{\pi}\right)^3 \,\approx \,0.00761 \, \textrm{meV} 
\end{displaymath}
and 0.00804\,meV for muonic deuterium.

However, I should remark that the contributions from figures\,1 and 2 of
Ref.\,\cite{kinoshita} were checked by direct integration.  Although the 
results agreed perfectly for the case of hydrogen, there were small, but
significant discrepancies for the case of deuterium. 
(hydrogen: Fig.\,1 contributes 0.000396\,meV and Fig.\,2 contributes
0.002931\,meV;  deuterium: direct integration gave 0.000472\,meV and
0.003364\,meV, respectively, while the work of ref.\cite{kinoshita}
indicates values  0.000419\,meV and 0.003906\,meV, respectively). 
This indicates that, at least for these two graphs, integration over 
momentum transfer involves more than a single reduced mass factor.


The hadronic vacuum polarization contribution has been estimated by a
number of authors \cite{hadron,friar99,eides}.  It amounts to about
0.012\,meV in hydrogen and 0.013\,meV in deuterium.  
One point that should not be forgotten about the hadronic VP correction
is the fact that the sum rule or dispersion relation that everyone
(including myself) used does not take into account the fact that the
proton  (nucleus) can 
in principle interact strongly with the hadrons in the virtual hadron
loop. It is irrelevant for the anomalous magnetic moment but probably not 
for muonic atoms.  An estimation of this effect appears to be extremely
difficult, and could easily change the correction by up to 50\%.  
Eides et al. \cite{eides} point out that the graph related to hadronic 
vacuum polarization can also contriibute to the measured value of the
nuclear charge distribution (and polarizability).  It is not easy to
determine where the contribution should be assigned.  This may also be
true for the so-called "proton self-energy" \cite{Pachuki2,eides}, which
involves some of the same graphs as are present in the calculation of
radiative corrections to electron scattering.

\subsubsection*{Finite nuclear size and nuclear polarization }
\vspace{-0.2cm}
The main contribution due to finite nuclear size has been given
analytically to order ($\alpha Z)^6$ by Friar \cite{friar79}.  The main
result is \\
\begin{equation}
\Delta E_{ns}\,=\,-\frac{2 \alpha Z}{3} \left(\frac{\alpha Z m_r}{n}\right)^3
 \cdot \left[\langle r^2 \rangle - \frac{\alpha Z m_r}{2} \langle r^3
 \rangle_{(2)} +(\alpha Z)^2 (F_{REL}+m^2_r F_{NR})  \right]
\label{eq:FS-friar}
\end{equation}
\noindent where $\langle r^2 \rangle$ is the mean square radius of the 
proton.  
For muonic hydrogen, the coefficient of $\langle r^2 \rangle$ is
5.1975\,(meV fm$^{-2}$), giving an energy shift (for the leading term)
of (3.979$\pm$0.076)\,meV if the proton rms radius is 0.875\,fm.  
Other values of the proton radius that have been reported recently in the
literature are 0.880\,fm \cite{rosenfelder} and (0.895\,$\pm$\,0.018\,fm)  
 \cite{sick}. 
The 
second term in Eq.(\ref{eq:FS-friar}) contributes 
-0.0232\,meV for a dipole form factor and  
-0.0212\,meV for a Gaussian form factor.  The parameters were fitted to
the proton rms radius.  This can be written as  
-0.0347$\langle r^2 \rangle^{3/2}$ or -0.0317$\langle r^2 \rangle^{3/2}$.
This differs slightly from the value given by Pachucki \cite{Pachuki2}.
The model dependence introduces an uncertainty about $\pm$0.002\,meV.   
The remaining terms contribute 0.00046\,meV.   This
estimate includes all of the terms given in \cite{friar79}, while other
authors \cite{Pachuki2} give only some of them.  Clearly the neglected
terms are not negligible. 
There is also a contribution of -$3\cdot10^{-6}$\,meV to the binding
energy of the $2p_{1/2}$-level, and a recoil correction of 
0.013\,meV to the binding energy of the 2s-level.   

 Pachucki \cite{Pachuki2} has estimated a correction similar to 
the second term (proportional to $\langle r^3 \rangle_{(2)} $) in
Eq.(\ref{eq:FS-friar}).  Since the logarithmic terms in the  
two-photon  correction without finite size (see below) also seem  to be
suspect,  this correction requires further investigation.  In
particular, the parametrization of the form factors used in any
calculation should reproduce the correct proton radius.    

For muonic deuterium, the main contribution amounts to \\
-6.0732\,$\langle r^2 \rangle$\,=\,-(27.787$\pm$0.078)\,meV.  
Depending on the model, the term proportional to $\langle r^3 \rangle_{(2)}$
gives a contribution of 0.382\,meV or 0.417\,meV.

As mentioned previously, the finite-size contributions to vacuum
polarization in muonic hydrogen can be parametrized as  
\mbox{$-\,0.0109 \langle r^2 \rangle \,-\, 0.0164 \langle r^2 \rangle$,}
giving a total of  $-0.0273\langle r^2
\rangle$ or -0.0209(6)\,meV if the proton radius is 0.875\,fm.  
For deuterium. only the contribution corresponding to the first term of
the sum \mbox{($-\,0.0129 \langle r^2 \rangle$)} has
been calculated.

The contribution due to nuclear polarization (in hydrogen) has been
calculated by  Rosenfelder \cite{rosenfelder99} to be
0.017\,$\pm$\,0.004\,meV, and by Pachuki \cite{Pachuki2} to be
0.012\,$\pm$\,0.002\,meV.  Other calculations \cite{srartsev,faustov}
give intermediate values 
(0.013\,meV and 0.016\,meV, respectively).  The value appearing in 
 table \ref{tab:final} is an average of the three most recent values,
 with the largest quoted uncertainty, which is probably underestimated.

\subsubsection*{Relativistic Recoil }
\vspace{-0.2cm}
As is well-known, the center-of-mass motion can be separated exactly
from the relative motion only in the nonrelativistic limit.
Relativistic corrections have been studied by many authors, and will not
be reviewed here.  The relativistic recoil corrections summarized in
\cite{RMP} include the effect of finite nuclear size to leading order in
$m_{\mu}/m_N$ properly.

Up to now this method has  been used to treat recoil corrections to
vacuum polarization only in the context of extensive numerical
calculations that include the Uehling potential in the complete
potential,  as described in \cite{RMP}.   They can be included
explicitly, as a perturbation correction to point-Coulomb values.  
Recall that (to leading order in $1/m_N$), the energy levels are given by 
\begin{equation}
  E~=~E_r - \frac{B_0^2}{2  m_N} + \frac{1}{2 m_N} \langle h(r) +  
  2 B_0 P_1(r) \rangle
\label{eq:rmp1}
\end{equation}
where $E_r$ is the energy level calculated using the reduced mass and 
$B_0$ is the unperturbed binding energy.  Also
\begin{equation}
   h(r)\, = \, - P_1(r)(P_1(r) + \frac{1}{r} Q_2(r))  
            - \frac{1}{3 r} Q_2(r) [P_1(r) + Q_4(r)/r^3]
\label{eq:rmp2}
\end{equation}
Here 
\begin{align} \label{eq:rmp3}
 P_1(r)&\,=\, 4 \pi \alpha Z \int_r^{\infty} r' \rho(r') dr' 
     & \,=\,&  -V(r)-rV'(r)  \\ \nonumber
Q_2(r)&\,=\,4 \pi \alpha Z \int_0^r r'^2 \rho(r') dr' & \,=\,&  r^2V'(r) \\ \nonumber   
Q_4(r)&\,=\,4 \pi \alpha Z \int_0^r r'^4 \rho(r') dr'   &     &    
\end{align}

An effective charge density $\rho_{VP}$ for vacuum
polarization  can be derived from the Fourier transform of the Uehling
potential. Recall that (for a point nucleus)
\begin{equation*} \begin{split}
V_{Uehl}(r) & \,=\,-\frac{\alpha Z}{r} \frac{2 \alpha}{3\pi} \cdot 
\chi_1(2 m_e r) \\ & \,=\, -(\alpha Z) \frac{2 \alpha}{3\pi} \cdot 
\int_1^{\infty} dz \frac{(z^2-1)^{1/2}}{z^2} \cdot 
\left(1+\frac{1}{2 z^2}\right) \left( \frac{2}{\pi} 
  \int_0^{\infty} \frac{q^2  \cdot  j_0(qr)}{q^2 + 4 m_e^2 z^2}\,dq\right) 
\end{split} \end{equation*}
\noindent where $\chi_n(x)$ is defined in \cite{RMP}.
In momentum space, the Fourier transform of 
$\nabla^2 V$ is obtained by multiplying the Fourier transform of $V$ by
$-q^2$.  
  Note that using the normalizations of \cite{RMP,hyperfine},
one has $\nabla^2 V = - 4 \pi \alpha Z \rho$ where $\rho$ is the 
charge density. 
One then obtains 
\begin{equation} \begin{split}
4 \pi \rho_{VP}(r) & \,=\, \frac{2 \alpha}{3\pi} \cdot 
\int_1^{\infty} dz \frac{(z^2-1)^{1/2}}{z^2} \cdot 
\left(1+\frac{1}{2 z^2}\right) \left(\frac{2}{\pi} 
\cdot \int_0^{\infty} \frac{q^4 \cdot j_0(qr)}{q^2 +4 m_e^2 z^2}\,dq \right) \\
& \,=\, \frac{2}{\pi} \cdot \int_0^{\infty} q^2 U_2(q)  j_0(qr)  \,dq
\label{eq:rhovp}
\end{split} \end{equation}
\noindent  $U_2(q)$ is defined in \cite{RMP}.   

Keeping only the Coulomb and Uehling potentials, one finds 
\begin{align*}
 P_1(r) &\,=\,- \alpha Z \frac{2 \alpha}{3\pi} (2 m_e) \chi_0(2 m_e r) \\
 Q_2(r) & \,=\, \alpha Z \left(1 + \frac{2 \alpha}{3\pi}[\chi_1(2 m_e r)
 +  (2 m_e r) \chi_0(2 m_e r)] \right)  \\
 Q_4(r) &\,=\,  \alpha Z \frac{2 \alpha}{3\pi} 
\int_1^{\infty} dz \frac{(z^2-1)^{1/2}}{z^2}  
\left(1+\frac{1}{2 z^2}\right) \\   & \cdot \left( \frac{2}{\pi} \right) 
 \int_0^{\infty} \frac{1}{q^2 + 4 m_e^2 z^2} \frac{(6qr-(qr)^3)\cos(qr)
   + (3(qr)^2-6)\sin(qr)}{q}  \,dq 
\end{align*}
\noindent 
Details of the calculations for the case of vacuum polarization are
given in Appendix\,1 and in Ref.\cite{Borie05}.  
Corrections due
 to finite nuclear size can be  included when a model for the charge
 distribution is given.  This done by Friar \cite{friar79} (and
 confirmed independently for two different model charge distributions); the
 contribution due to finite nuclear size to the recoil correction for
 the binding energy of the 2s-level  is -0.013\,meV.  The factor $1/m_n$
 is replaced by  $1/(m_{\mu}+m_N)$, also consistent with the
 calculations presented in \cite{friar79}.  

Combining the expectation values given in Appendix\,1 according to 
equations \ref{eq:rmp1}
and \ref{eq:rmp2}, one finds a contribution to the 2p-2s transition of 
\mbox{-0.00419\,meV} (hydrogen) and \mbox{-0.00479\,meV} (deuterium).  
To obtain the full relativistic and recoil corrections,
one must add the difference between the expectation values of the
Uehling potential calculated with relativistic and nonrelativistic wave
functions, 
giving a total correction of 0.0166\,meV for muonic hydrogen.  This is
in quite good agreement with the correction of .0169\,meV calculated by 
Veitia and Pachucki \cite{Pachuki3}.  The treatment presented here has the
advantage of avoiding second order perturbation theory.
For deuterium, one obtains a total correction of 0.0179\,meV.

The review by Eides et al. \cite{eides} gives a better version of the two
photon recoil (Eq. 136) than was available for the review by Borie and
G. Rinker \cite{RMP}.  Evaluating this expression for muonic hydrogen
gives a contribution of -0.04497\,meV to the 2p-2s transition in
hydrogen and -0.02656\,meV in deuterium.
Higher order radiative recoil corrections give an additional
contribution (in hydrogen) of -0.0096\,meV \cite{eides}. 
However, some of the contributions to the expressions given in
\cite{eides} involve logarithms of the mass ratio $m_{\mu}/m_N$.
Logarithms can only arise in integrations in the region from 
$m_{\mu}$ to $m_N$; in this region the effect of the nuclear form factor
should not be neglected.  Pachucki \cite{Pachuki1} has estimated a
finite size correction to this of about 0.02\,meV, which seems to be
similar to the term proportional to $\langle r^3 \rangle_{(2)} $ given
in Eq.(\ref{eq:FS-friar}) as calculated in the external field
approximation by Friar \cite{friar79}.  This two-photon correction
requires further investigation.  In particular, the parametrization of
the form factors used in any calculation should reproduce the correct
proton radius.   Also the relationship among the different contributions
needs to be specified more clearly.

An additional recoil correction for states with $\ell \, \ne \, 0$ has
been given by \cite{barker} (see also \cite{eides}).  It is 
\begin{equation}
\Delta E_{n, \ell ,j}~=~\frac{(\alpha Z)^4 \cdot m_r^3}{2 n^3 m_N^2} 
 (1-\delta_{\ell 0}) \left(\frac{1}{\kappa (2\ell+1)} \right)
\label{eq:recoil1}
\end{equation}
When evaluated for the 2p-states of muonic hydrogen, one finds a
contribution to the 2p-2s transition energy of 0.0575\,meV for the
2p$_{1/2}$ state and -0.0287\,meV for the 2p$_{3/2}$ state in hydrogen
( 0.0168\,meV for the
2p$_{1/2}$ state and -0.0084\,meV for the 2p$_{3/2}$ state in deuterium)

A final point about recoil corrections is that in the case of light
muonic atoms, the mass ratio $m_{\mu}/m_N$ is considerably larger than
the usual perturbation expansion parameter $\alpha Z$.  Contributions of
higher order in the mass ratio could be significant.

\subsubsection*{Muon Lamb Shift }
\vspace{-0.2cm}
For the calculation of muon self-energy and vacuum polarization, the
lowest order (one-loop approximation) contribution is well-known, at
least in perturbation theory.
Including also  muon vacuum polarization
(0.0168\,meV) and an extra term of order $(Z \alpha)^5$ as given in
\cite{eides}:
which contributes -0.00443\,meV, one finds a  contribution of
-0.66788\,meV for the \mbox{$2s_{1/2}-2p_{1/2}$} transition and 
 -0.65031\,meV for the  \mbox{$2s_{1/2}-2p_{3/2}$} transition.  
For deuterium, the corresponding contributions are given by 
-0.77462\,meV for the \mbox{$2s_{1/2}-2p_{1/2}$} transition and 
 -0.75512\,meV for the  \mbox{$2s_{1/2}-2p_{3/2}$} transition.  
The second order calculation in deuterium includes muonic vacuum 
polarization (0.01968\,meV); the extra term of order $(Z \alpha)^5$ as 
given in \cite{eides},  contributes -0.00518\,meV. 

These results, and the higher order corrections \cite{RMP,HelvPA} 
can be summarized as \\
\begin{table}[!h]
\begin{center} 
    \begin{tabular}{|lrr|}
      \hline
Transition   &  $2p_{1/2}-2s_{1/2}$  & $2p_{3/2}-2s_{1/2}$            \\
\hline
Hydrogen   &       &     \\           
second order          &  -0.66788   &    -0.65031            \\
higher orders        &   -0.00172    &   -0.00165        \\
         &      &     \\   
Total     &     -0.66960   &    -0.65196   \\
\hline
\hline
Deuterium   &       &     \\           
second order          &   -0.774616 &    -0.755125            \\
higher orders        &   -0.002001     &   -0.001926        \\
      &         &          \\
Total     &     -0.776617   &    -0.757051   \\
\hline
\end{tabular}        
 \caption{ Contributions to the muon Lamb shift  
($ E(2p_{1/2}) - E(2s_{1/2})$)  in muonic  hydrogen and deuterium, in meV.  }
\end{center}   
\end{table}



For hydrogen, Pachuki \cite{Pachuki1} has estimated an additional
contribution of -0.005\,meV for a contribution corresponding to a vacuum
polarization insert in the external photon. 

The higher order contributions can be written in the form
\begin{displaymath}
\Delta E_{LS}\,=\, \frac{1}{m_{\mu}^2} \cdot \langle\nabla^2 V\rangle
 \big[m_{\mu}^2 F'_1(0) + \frac{a_{\mu}}{2} \big] + \frac{a_{\mu}}{2}m_{\mu}^2
\Big\langle \frac{2}{r}\,\frac{dV}{dr} \vec{L} \cdot \vec{\sigma}_{\mu}
 \Big\rangle 
\end{displaymath}
where $F_2(0) \,=\,a_{\mu}$; the higher order contributions (fourth and
sixth) can be taken from the well-known theory of the muon's anomalous
magnetic moment: \\ \mbox{$F_2(0) \,=\,a_{\mu}=\alpha/2 \pi +
  0.7658(\alpha/\pi)^2 + 24.05(\alpha/\pi)^3$}.  \\
  The fourth order contribution to $F'_1(0)$ is \\ 
$0.46994(\alpha/\pi)^2 + 2.21656(\alpha/\pi)^2\,=\,2.68650(\alpha/\pi)^2$
\cite{RMP}.  The sixth order contributions to  $F'_1(0)$ that involve 
electron vacuum polarization loops (especially the light-by-light graph)
might contribute at an experimentally significant level, but have not
been calculated.

\newpage
\subsubsection*{Summary of contributions for muonic hydrogen}
\vspace{-0.2cm}
Using the fundamental constants from the CODATA 2002 (\cite{codat02})  
one finds the  transition energies in meV in table \ref{tab:final}.  Here
the main  vacuum polarization contributions are given for a   
point nucleus, using the Dirac equation with reduced mass. 
  Some uncertainties have been increased from the values
given by the authors, as discussed in the text. 

The finite size corrections for hydrogen up to order $(\alpha Z)^5$ 
can be parametrized as \\
 \mbox{$ 5.1975 \langle r^2 \rangle \,+\,0.0109 \langle
 r^2 \rangle\,+\, 0.0164 \langle r^2 \rangle\,+\,0.0347 \langle r^3
 \rangle_{(2)}$.}  The various contributions are discussed in the text. 
\begin{table}[!h]
\begin{center} 
    \begin{tabular}{|lrr|}
      \hline
Contribution  &  Value (meV) &  Uncertainty (meV) \\
      \hline
Uehling        &     205.0282~    &  \\
K\"allen-Sabry     &       1.5081~  &           \\
Wichmann-Kroll   &      -0.00103     & \\
virt. Delbrueck    &     0.00135   &  0.00015\\
mixed mu-e VP    &       0.00007   &  \\
hadronic VP       &      0.011~~~    &   0.002~~~     \\       
sixth order \cite{kinoshita}  & 0.00761    &   \\
\hline
recoil  \cite{eides} (eq136)     & -0.04497  &  \\
recoil, higher order \cite{eides} & -0.0096~ & \\
recoil, finite size \cite{friar79} & 0.013~~~  &  0.001~~~ \\
recoil correction  to VP \cite{RMP}  &  -0.0042~~ &  \\
additional recoil  \cite{barker} & 0.0575~~ &   \\
\hline
muon Lamb shift   &    &  \\
second order      &     -0.66788 & \\
fourth order      &     -0.00169 &  \\
\hline
nuclear size  ($R_p$=0.875\,fm)  & &  0.007\,fm \\
main correction \cite{friar79}  & -3.979~~    &  0.076~~~   \\
order $(\alpha Z)^5$ \cite{friar79} &  0.0232~  &  0.002~~~ \\
order $(\alpha Z)^6$ \cite{friar79} &  -0.0005~  &  \\
correction to VP  &   -0.0083~  &  \\
polarization      &    0.015~~~ &  0.004~~          \\
\hline
Other (not checked)  &    &  \\
VP iterations \cite{Pachuki1}  &  0.151~~~  &    \\
VP insertion in self energy \cite{Pachuki1}  &  -0.005~~~  &    \\ 
additional size for VP   \cite{eides}  &  -0.0128~~  &    \\ 
\hline
\end{tabular}        
 \caption{ Contributions to the muonic hydrogen Lamb shift
 (the $2s_{1/2}-2p_{1/2}$ transition). The proton
   radius is taken from \cite{codat02}. }
  \label{tab:final}
\end{center}   
\end{table}        


\subsubsection*{Summary of contributions for muonic deuterium }
\vspace{-0.2cm}
For deuterium, 
one finds the  transition energies in meV in table \ref{tab:Total-D}.  
Also here the main  vacuum polarization contributions are given for a   
point nucleus, using the Dirac equation with reduced mass. 
The finite size corrections for deuterium up to order $(\alpha Z)^5$ can be
parametrized as \\   
 \mbox{$ 6.0732 \langle r^2 \rangle \,+\,0.0129 \langle
 r^2 \rangle\, +\,0.0409 \langle r^3
 \rangle_{(2)}$}, although not all contributions to the effect of
 finite size on the vacuum polarization correction are included.  
\begin{table}[!h]
\begin{center} 
    \begin{tabular}{|lrr|}
      \hline
Contribution  &  Value (meV) &  Uncertainty (meV) \\
      \hline
Uehling        &     227.6577~    &  \\
K\"allen-Sabry     &       1.6662~  &           \\
Wichmann-Kroll   &      -0.00111     & \\
virt. Delbrueck    &     0.00147   &  0.00016\\
mixed mu-e VP    &       0.00008   &  \\
hadronic VP       &      0.013~~~    &   0.002~~~     \\       
sixth order \cite{kinoshita}  & 0.00804    &   \\
\hline
recoil  \cite{eides} (eq136)     & -0.02656  &  \\
recoil, higher order \cite{eides} & ?~~~ & \\
recoil, finite size \cite{friar79} & 0.019~~~  &  0.003~~~ \\
recoil correction  to VP \cite{RMP}  &  -0.0048~~ &  \\
additional recoil  \cite{barker} & 0.0168~~ &   \\
\hline
muon Lamb shift   &    &  \\
second order      &     -0.77462 & \\
fourth order      &     -0.00200 &  \\
\hline
nuclear size  ($R_d$=2.139\,fm)  & &  0.003\,fm \\
main correction \cite{friar79}  & -27.787~~    &  0.078~~~   \\
order $(\alpha Z)^5$ \cite{friar79} &  0.0400~  &  0.018~~~ \\
order $(\alpha Z)^6$ \cite{friar79} &  -0.0045~  &  \\
correction to VP  &   -0.0592~  &  \\
polarization      &    ?~~~ &  ~~          \\
\hline
Other (not checked)  &    &  \\
VP iterations \cite{Pachuki1}  &  ?~~~  &    \\
VP insertion in self energy \cite{Pachuki1}  &  ?~~~  &    \\ 
additional size for VP   \cite{eides}  &  ?~~~  &    \\ 
\hline
\end{tabular}        
 \caption{ Contributions to the muonic deuterium Lamb shift. The 
 deuteron  radius is taken from \cite{codat02}. }
  \label{tab:Total-D}
\end{center}   
\end{table}

\subsubsection*{Fine structure of the 2p state }
\vspace{-0.2cm}
The fine structure of the 2p states can be calculated  by using
the relativistic Dirac energies, computing the vacuum polarization
contributions with Dirac wave functions, and including the effect of the
anomalous magnetic moment in the muon Lamb shift.  
An additional recoil correction (Eq.\ref{eq:recoil1}) also has to be
included. 
The results are summarized in table \ref{tab:FS}. 
\begin{table}[!h]
\begin{center} 
    \begin{tabular}{|lrr|}
      \hline
          &  Hydrogen  &   Deuterium  \\
Dirac          &      8.41564   &  8.86430            \\
Uehling(VP)        &     0.0050~    &  0.00575  \\
K\"allen-Sabry      &     0.00004  &   0.00005    \\
\hline
anomalous moment $a_{\mu}  $ &     &     \\
second order      &      0.01757   &  0.01491  \\
higher orders      &     0.00007   &  0.00007  \\
\hline
Recoil  (Eq.(\ref{eq:recoil1}))     &  -0.0862~  & -0.0252~    \\
\hline
Total Fine Structure    &      8.352~~  &  8.864~~  \\
\hline
\end{tabular}        
 \caption{ Contributions to the fine structure 
($ E(2p_{3/2}) - E(2p_{1/2})$) of the 2p-state in muonic
   hydrogen and deuterium, in meV.  }
  \label{tab:FS}
\end{center}   
\end{table} 
 One should also include the 
$B^2/2M_N$-type correction to the fine structure. \\ (see \cite{eides},
Eq(38)).  This is tiny 
($5.7 \cdot 10^{-6}$\,meV in hydrogen) and is not included in the
table.  Friar \cite{friar79} has given  expressions for the energy
shifts of the 2p-states due to finite nuclear size.  
These were calculated and found to give a negligible contribution ($3.1
\cdot 10^{-6}$\,meV) to the fine structure of the 2p-state in hydrogen.

\subsection*{Hyperfine structure }
\vspace{-0.2cm}
The Breit equation  \cite{barker,eides,hyperfine} contributions to the
fine- and hyperfine interactions for general potentials and 
arbitrary spins were given by Metzner and Pilkuhn \cite{rmetzner}.  Here 
a version applicable to the case of muonic atoms ($Z_1=-1$, $s_1=1/2$,
$m_1=m_{\mu}$,  $\kappa_1=a_{\mu}$, $Z_2=Z$) is given. 
\begin{equation}
 V_{L,s_1} =  \frac{1}{2 m_{\mu}}\, \frac{1}{r}\,\frac{dV}{dr} 
\Bigl[ \frac{1+a_{\mu}}{s_1 m_r} - \frac{1}{m_{\mu}} \Bigr]
 \vec{L} \cdot \vec{s}_{1}
\label{eq:fs1}
\end{equation}%
This can be rearranged to give the well-known form for spin 1/2
particles with an anomalous magnetic moment, namely  
\begin{equation*}
- \frac{1}{r}\,\frac{dV}{dr} \cdot \frac{1+a_{\mu}+(a_{\mu}+1/2)m_N/m_{\mu}}
{m_N m_{\mu}} \vec{L} \cdot \vec{\sigma}_{\mu}
\end{equation*}%
Note that 
\begin{displaymath}
 \frac{1}{m_N m_{\mu}}+\frac{1}{2 m_{\mu}^2}\,=\,
  \frac{1}{2 m_r^2}-\frac{1}{2 m_N^2}
\end{displaymath}
so that the terms not involving  $a_{\mu}$ in the spin-orbit
contribution are really the Dirac fine
structure plus the Barker-Glover correction (Eq. \ref{eq:recoil1}).
 
Also
\begin{equation*}
 V_{L,s_2} =  \frac{1}{2 m_{2}}\, \frac{1}{r}\,\frac{dV}{dr} 
\Bigl[ \frac{1+\kappa_{2}/Z}{s_2 m_r} - \frac{1}{m_{2}} \Bigr]
 \vec{L} \cdot \vec{s}_{2}
\end{equation*}%
Usually one writes 
\begin{equation*}
  \frac{Z+\kappa_{2}}{ m_2} = \frac{\mu_2}{m_{p}} 
\end{equation*}%
\noindent   where $\mu_2$ is the magnetic moment of the nucleus in units
of nuclear magnetons \mbox{($\mu_N=e/2 m_p$).}  A value of 
$\mu_d\,=\,0.85744\,\mu_N\,=\,0.307012\,\mu_p$ corresponds to
$\kappa_d$\,=\,0.714. 
\begin{equation*}
 V_{s_1,s_2} = \frac{2 (1+a_{\mu})\mu_2}{2 s_2 m_{\mu} m_{2}}\, 
 \Bigl[ \frac{1}{r}\,\frac{dV}{dr} 
(3 \vec{s}_1 \cdot \hat{r} \vec{s}_2 \cdot \hat{r} - \vec{s}_1
 \cdot \vec{s}_{2}) - \frac{2}{3} \nabla^2 V \vec{s}_1 \cdot \vec{s}_{2}
 \Bigr]  
\end{equation*}%
\begin{equation*}
 V_{Q} = - \alpha Q \,  \frac{1}{r}\,\frac{dV}{dr} 
\Bigl[3 \vec{s}_2 \cdot \hat{r} \vec{s}_2 \cdot \hat{r} - \vec{s}_2
 \cdot \vec{s}_{2} \Bigr] 
\end{equation*}%
\noindent  with Q in units of $1/m_2^2$.  
 The quadrupole moment of the deuteron is taken to be
Q\,=\,0.2860(15)\,fm$^2$~\cite{friar02,reid-v,bishop}.  In other units,
one finds  Q\,=\,25.84/$m_d^2$\,=\,7.345$\times10^{-6}$\,MeV$^{-2}$.\\

Note that $V_{L,s_1}$ describes the fine structure, while the hyperfine
structure is described (in perturbation theory) by the expectation values
of $V_{L,s_2}$, $V_{s_1,s_2}$, and $V_{Q}$ (where applicable). 

%

The Uehling potential has to be included in the potential $V(r)$.  
For states with $\ell\,>\,0$ in light atoms, and
neglecting the effect of finite nuclear size, we may take 
\begin{equation}
\frac{1}{r} \frac{d V}{dr}~=~\frac{ \alpha Z}{r^3} \cdot \left[ 1 +
 \frac{2 \alpha}{3 \pi} \int_1^{\infty} \frac{(z^2-1)^{1/2}}{z^2}\cdot
\left(1+\frac{1}{2 z^2}\right) \cdot (1 + 2 m_e r z) \cdot e^{-2 m_e r z} \,dz \right]
\label{eq:vp2p}
\end{equation}
which is obtained from the Uehling potential \cite{Uehling,serber}
by differentiation.  Then, assuming that it is sufficient to use
nonrelativistic point Coulomb wave functions for the 2p state, one finds 
\begin{displaymath}
\Big\langle \frac{1}{r^3} \Big\rangle_{2p} \rightarrow 
 \Big\langle \frac{1}{r^3} \Big\rangle_{2p} \cdot (1+\varepsilon_{2p})
\end{displaymath}
\noindent  where 
\begin{equation}
\varepsilon_{2p} ~=~ \frac{2 \alpha}{3 \pi} 
 \int_1^{\infty} \frac{(z^2-1)^{1/2}}{z^2}\cdot \left(1+\frac{1}{2 z^2}\right)
  \cdot \left(\frac{1}{(1+az)^2} + \frac{2 az}{(1+az)^3}\right) \,dz
\label{eq:eps2p}
\end{equation}
\noindent  with $a \,=\,2 m_e /(\alpha Z m_r)$.  
For hydrogen,  $\varepsilon_{2p}$\,=\,0.000365, and for deuterium 
$\varepsilon_{2p}$\,=\,0.000391.


\subsubsection*{Hyperfine structure of the  2p state in muonic hydrogen}
\vspace{-0.2cm}
The hyperfine structure of muonic hydrogen is calculated in the same way
as was done in 
earlier work \cite{hyperfine,BorieHe3}, but with improved accuracy.  
Most of the formalism and results are similar to those given by
\cite{Pachuki1} and \cite{brodsky-p}. \\

%
The hyperfine structure of the 2p-state is given by \cite{hyperfine,brodsky-p}
($F$ is the total angular momentum of the state)
\begin{equation} \begin{split}
\frac{1}{4m_{\mu} m_N} \Big\langle \frac{1}{r}\,\frac{dV}{dr}\Big\rangle_{2p}
 & \cdot(1+\kappa) \biggl[2(1+x_p) \delta_{j j'} (F(F+1)-11/4) \\
 & + 6 \hat{j} \hat{j}' (C_{F1}(1+a_{\mu})-2(1+x)) \left\{ \begin{array}{ccc} 
 \ell & F & 1 \\
 \frac{1}{2} & \frac{1}{2} & j 
 \end{array} \right\}  \left\{  \begin{array}{ccc} 
 \ell & F & 1 \\
 \frac{1}{2} & \frac{1}{2} & j' 
 \end{array} \right\}     \biggr] 
\end{split}
\end{equation}

\noindent where $\hat{j} = \sqrt{2 j + 1}$, the 6-j symbols are defined
 in \cite{edmonds},  and \\
 $C_{F1}=\delta_{F1}-2\delta_{F0}-(1/5)\delta_{F2}$ 


Also
\begin{displaymath}
 x_p\,=\, \frac{m_{\mu} (1 + 2 \kappa_p)}{2 m_p (1 + \kappa_p)}\,=\,0.09245
\end{displaymath}
\noindent  represents a recoil correction due to Thomas precession
\cite{hyperfine,barker,brodsky-p}.
The correction due to vacuum polarization (Eq.\,(\ref{eq:eps2p})) 
should be applied to the HFS shifts of the 2p-states.  \\ 

As has been known for a long time \cite{hyperfine,BorieHe3,Pachuki1,brodsky-p},
the states with total angular momentum $F=1$ are a superposition of the
states with $j=1/2$ and $j=3/2$.  
Let the fine structure splitting be denoted by 
$\delta \,=\, E_{2p3/2} - E_{2p1/2}$\,=\,8.352\,meV, and let 
\begin{displaymath}
\beta_p\,=\,\frac{(\alpha Z)^4 m_r^3}{3m_{\mu} m_p}
 \cdot (1 + \kappa_p) 
\end{displaymath}
\noindent  and $\beta ' \,=\, \beta_p \cdot (1+ \varepsilon_{2p})$.
The matrix elements for the hyperfine structure of the 2p-state 
are then given by \\
\begin{center} 
    \begin{tabular}{ccl}
 $j$  &    $j'$      &   Energy           \\
     \hline
1/2   & 1/2   &  $(\beta'/8) (2+x_p+a_{\mu})[ -\delta_{F,0} + 1/3
    \, \delta_{F,1}]   $           \\
3/2  &  3/2   & $\delta\,+\,(\beta'/4)(4+5x_d -a_{\mu}) [-1/12 \,
     \delta_{F,1}  + 1/20 \,\delta_{F,2} ]$     \\
3/2  & 1/2  & $(\beta'/24) (1 + 2x_p -a_{\mu}) [\sqrt{2}\,\delta_{F,1}]$
     \\
     \hline
\end{tabular}        
\end{center}

Then for the 2p-level with $j=j'=1/2$ and $F=0$, the energy shift is
given by
\mbox{$-\,(\beta'/8)(2+x_p+a_{\mu})$\,=\,-5.971\,meV,}      
and for the 2p-level with $j=j'=3/2$ and $F=2$, the energy shift is
given by
\mbox{$\delta\,+\,(\beta'/80)(4+5x_p-a_{\mu})$\,=\,9.6243\,meV.}    


For the 2p-levels with  $F=1$  the corresponding matrix 
has to be diagonalized.    The resulting numerical values for the 
eigenvalues are  $(\Delta \pm R)/2$\,=\,1.846\,meV and 6.376\,meV,  
where
\begin{displaymath}
\Delta\,=\, \delta - \beta '(x_p - a_{\mu}) /16
\end{displaymath}
\begin{displaymath}
 R^2 \,=\, [\delta  - \beta '(1 +7x_p/8 + a_{\mu}/8) /6]^2
 + (\beta')^2 (1+2x_p-a_{\mu})^2/288  
\end{displaymath}


\subsubsection*{Hyperfine structure of the 2p-state in muonic deuterium}
\vspace{-0.2cm}
For the 2p state, the matrix elements of the magnetic hyperfine
structure have been given by Brodsky and Parsons \cite{brodsky-p}.    
For hydrogen they are the same as those calculated above. Here the
Uehling potential will be included in the expectation value of 
\begin{equation*}
  \Big\langle \frac{1}{r}\,\frac{dV}{dr} \Big\rangle
\end{equation*}%
as discussed above.  

Let 
\begin{equation*}
 \beta_D\,=\, \frac{16 (1+\kappa_d)}{m_{\mu} m_{d}}  \,\frac{\alpha}
 {(\alpha Z m_r/n)^3}{\ell(\ell+1)(2\ell+1)} \,=\, \frac{(1+\kappa_d)}{6
 m_{\mu} m_{d}} (\alpha Z m_r)^3 \,=\,4.0906\,meV     
\end{equation*}%
(for a point Coulomb potential)


The matrix elements for the magnetic hyperfine structure are then given
by \\
\begin{center} 
    \begin{tabular}{ccl}
 $j$  &    $j'$      &   Energy           \\
     \hline
1/2   & 1/2   &  $(\beta_D/6) (2+x_d+a_{\mu})[ -\delta_{F,1/2} + 1/2
    \, \delta_{F,3/2}]   $           \\
3/2  &  3/2   & $\delta\,+\,(\beta_D/4)(4+5x_d -a_{\mu}) [-1/6 \,
     \delta_{F,1/2} - 1/15 \,\delta_{F,3/2} + 1/10 \,\delta_{F,5/2} ]$     \\
3/2  & 1/2  &  $(\beta_D/48) (1 + 2x_d -a_{\mu}) [\sqrt{2} \,\delta_{F,1/2}  
   - \sqrt{5}\, \delta_{F,3/2} ]$             \\
     \hline
\end{tabular}        
\end{center}   
where $x_d\,=\,(m^2_{\mu}/ m_{d} m_r)(\kappa_d/(1+\kappa_d))$\,=\,0.0248.

For the evaluation of the contributions of the quadrupole HFS, let 
\begin{equation*}
 \epsilon_Q =  \alpha Q \, \Big\langle \frac{1}{r}\,\frac{dV}{dr} \Big\rangle
\end{equation*}%

For a point Coulomb potential, and the 2p-state, 
\mbox{$\epsilon_Q\,=\,\alpha Q (Z \alpha m_r)^3/24 \,=\,$0.43243\,meV.}  
The quadrupole interaction results in energy shifts of \\
\begin{center} 
    \begin{tabular}{ccl}
 $j$  &    $j'$      &   Energy           \\
     \hline
1/2   & 1/2   &  $0$           \\
3/2  &  3/2   & $ \epsilon_Q\,[\delta_{F,1/2} - 4/5 \,\delta_{F,3/2} + 
   1/5 \, \delta_{F,5/2} ]$~                \\
3/2  & 1/2  &  $ \epsilon_Q\,[ \sqrt{2} \, \delta_{F,1/2} - 
   1/\sqrt{5} \,\delta_{F,3/2} ]$             \\
     \hline
\end{tabular}        
\end{center}   
  
As mentioned before, the Uehling potential has to be included in the
potential $V(r)$.  For states with  $\ell\,>\,0$ in light atoms, this
can be taken into account by multiplying $\beta_D$ and $\epsilon_Q$ by  
(1+ $\varepsilon_{2p}$) where $\varepsilon_{2p}$ is given by
Eq.(\ref{eq:eps2p}). 
With a numerical value of $\varepsilon_{2p}$\,=\,0.000391 for muonic
deuterium, the value of $\epsilon_Q$ is increased to 0.43440\,meV and
the value of $\beta_D$ is increased to $\beta'_D$=4.0922\,meV.    

Then for the 2p-level with $j=j'=3/2$ and $F=5/2$, the energy shift is
given by
\mbox{$\delta\,+\,\epsilon_Q/5\,+\,(\beta'_D/40)(4+5x_d-a_{\mu})$\,=\,9.373\,meV.}    


For the 2p-levels with  $F=1/2$ and  $F=3/2$, the corresponding matrices
have to be diagonalized.    The resulting numerical values for the 
eigenvalues are, \\ for  $F=1/2$, -1.3834\,meV and 8.5974\,meV;  
for  $F=3/2$ they are 0.6856\,meV and  8.2410\,meV.   


\subsubsection*{Hyperfine structure of the  2s-state:}
\vspace{-0.2cm}
The expectation value of $V_{s_1 s_2}$ in an ns state with $j=1/2$ is 
\begin{displaymath}
\Delta E_{ns}=\,=\,\frac{2 \mu_2 \alpha (\alpha Z)^3 m_r^3}
{3 n^3 m_{\mu} m_2 s_2} \cdot (1 + a_{\mu})  [F(F+1)-s_2(s_2+1)-3/4]
\end{displaymath}
When $s_2=1/2$, and $\mu_2/m_p=(1+\kappa_2)/m_2$, this reproduces the
well-known result for muonic hydrogen: 
\begin{displaymath}
\Delta  E_{ns}\,=\,\frac{8 (\alpha Z)^4 m_r^3}{3 n^3 m_{\mu} m_2}
 \cdot (1 + \kappa_2) \cdot (1 + a_{\mu}) \,=\,(8/n^3) \beta_p 
\cdot (1 + a_{\mu})  \,=\,  (8/n^3) \times 22.8332\,meV 
\end{displaymath}
\noindent
(see, for example \cite{eides}, Eq. (271,277)).  The numerical value
was calculated for hydrogen. 
For deuterium, with $s_2=1$, the corresponding hyperfine splitting is 
\begin{displaymath}
\Delta  E_{ns}\,=\,\frac{2 (\alpha Z)^4 m_r^3}{3 n^3 m_{\mu} m_2}
 \cdot (1 + \kappa_d) \cdot (1 + a_{\mu}) \cdot [F(F+1) - 11/4] 
 \,=\,  (8/n^3) \times 2.04766\,meV \times [F(F+1) - 11/4] 
\end{displaymath}
\noindent for a total splitting of 6.14298\,meV in muonic deuterium.  
This is in reasonably good agreement with the result given by 
Carboni \cite{carboni}.


As was shown in \cite{hyperfine,eides}, the energy shift of the 2s-state 
in muonic hydrogen is given by: 
\begin{equation}
\Delta E_{2s}~=~\beta \cdot (1 + a_{\mu}) \cdot (1+\varepsilon_{VP} +
 \varepsilon_{vertex} + \varepsilon_{Breit} + \varepsilon_{FS,rec}) 
 \cdot [\delta_{F1}-3\delta_{F0}]/4  
\label{eq:hf2s}
\end{equation}
\noindent The corrections due to QED effects, nuclear size and recoil
 are analogous for muonic deuterium.      

The QED corrections have been  discussed by Borie
\cite{Borie05,hyperfine,BorieHe3} (see also \cite{Brodsky}),  and are
given in Appendix\,2.   

The correction due to finite size and recoil have been given in
\cite{Pachuki1} as -0.145\,meV, while a value of -0.152\,meV is given in
\cite{martynenko}.   Ref.\,\cite{Pachuki1} also gives 
 a correction as calculated by Zemach
(\cite{zemach}) equal to -0.183\,meV, but claims that this correction
does not treat recoil properly.   The Zemach correction is equal to 
\begin{displaymath}
\varepsilon_{Zem} \,=\, -2  \alpha Z m_r \langle r \rangle_{(2)}
\end{displaymath}
where  $\langle r \rangle_{(2)}$ 
is given in \cite{hyperfine,friar79,friar04}.  Using the value 
$\langle r \rangle_{(2)}\,=\,1.086 \pm 0.012$\,fm from \cite{friar04},
gives   $\varepsilon_{Zem}\,=\,-0.00702$, and a contribution of 
of \mbox{-0.1742\,meV} to the hyperfine splitting of the 2s state.
Including this, but not other recoil corrections to the hyperfine
structure of the 2s-state gives a total splitting of 22.7806\,meV.
Additional higher order corrections calculated in
Ref.\,\cite{martynenko} amount to a total of -0.0003\,meV and are not
included here.  

It would be very desirable to understand the reasons for the discrepancy
between references \cite{Pachuki1} and \cite{martynenko} in the
calculations of this effect.  Also, since the Zemach radius seems to be
sensitive to details of the electric and magnetic charge distributions  
\cite{friar04}, evaluations performed with a dipole-type form factor may
not be good enough.  This point requires further invesigation.

For muonic deuterium, the coefficient of 
$\langle r \rangle_{(2)}$ is -0.007398\,fm$^{-1}$, giving, with  \\
\mbox{$\langle r \rangle_{(2)}\,=\,2.593 \pm 0.016$\,fm} from \cite{friar04},
$\varepsilon_{Zem}\,=\,-0.01918 \pm 0.00012$.

The total hyperfine splitting of the 2s-state of muonic deuterium,
including all corrections, is  
\begin{equation*}
\Delta E_{2s}~=~ \frac{3}{2} \beta_D \cdot (1 + a_{\mu}) \cdot
 (1+\varepsilon_{VP} + \varepsilon_{vertex} + \varepsilon_{Breit} +
 \varepsilon_{FS,rec}) \,=\,6.0582\,meV     
\end{equation*}

\newpage
\begin{table}[!h]
\begin{center} 
    \begin{tabular}{|lr|}
     \hline
Transition   &   Energy shift in meV  \\
     \hline
$^1 p_{1/2} - ^1 s_{1/2}$        &   11.114       \\
$^3 p_{1/2} - ^1 s_{1/2}$       &    18.931        \\
$^3 p_{3/2} - ^1 s_{1/2}$         &  23.461         \\
$^1 p_{1/2} - ^3 s_{1/2}$        &   -11.666       \\
$^3 p_{1/2} - ^3 s_{1/2}$       &      -3.849        \\
$^3 p_{3/2} - ^3 s_{1/2}$         &    0.681         \\
$^5 p_{3/2} - ^3 s_{1/2}$       & 3.929   \\
     \hline
\end{tabular}        
 \caption{ Fine- and hyperfine contributions to the Lamb shift in muonic
   hydrogen.   }
  \label{tab:HFS-mu-p}
\end{center}   
\end{table}    
\begin{table}[!h]
\begin{center} 
    \begin{tabular}{|lr|}
     \hline
Transition   &   Energy shift in meV  \\
     \hline
$^2 p_{1/2} - ^2 s_{1/2}$        &   2.655       \\
$^2 p_{3/2} - ^2 s_{1/2}$       &    12.636        \\
$^4 p_{1/2} - ^2 s_{1/2}$       &     4.724        \\
$^4 p_{3/2} - ^2 s_{1/2}$         &  12.280         \\
$^2 p_{1/2} - ^4 s_{1/2}$        &   -3.403       \\
$^2 p_{3/2} - ^4 s_{1/2}$         &    6.578         \\
$^4 p_{1/2} - ^4 s_{1/2}$       &      -1.334        \\
$^6 p_{3/2} - ^4 s_{1/2}$       & 6.222   \\
$^6 p_{3/2} - ^4 s_{1/2}$       & 7.354   \\
     \hline
\end{tabular}        
 \caption{ Fine- and hyperfine contributions to the Lamb shift in muonic
   deuterium.   }
  \label{tab:HFS-mu-d}
\end{center}   
\end{table}    
Tables \ref{tab:HFS-mu-p} and \ref{tab:HFS-mu-d} give the contributions
to the transition energies due to fine and hyperfine structure.

\subsubsection*{Summary of contributions and Conclusions}
\vspace{-0.2cm}
 The most important contributions to the Lamb shift in muonic hydrogen,
including hyperfine structure, have been independently recalculated. A new
calculation of  some terms that were
omitted in the most recent literature, such as the virtual Delbr\"uck 
effect \cite{Borie76} and an alternative calculation of the relativistic
recoil correction have been  presented. 

Numerically the results given in  table \ref{tab:final} add up to a
total correction of \\
 \mbox{(206.032(6)  - 5.225\,$\langle r^2 \rangle$ 
+ 0.0347\,$\langle r^2 \rangle^{3/2}$)\,meV\,=\,202.055$\pm$0.12\,meV.}    
(for the value of the proton radius from \cite{codat02}).
As is well known, most of the uncertainty arises from the uncertainty in 
the proton radius. 

Numerical results were also given for muonic deuterium.
The total correction is \\
 \mbox{(228.573(6)  - 6.086\,$\langle r^2 \rangle$ 
+ 0.0409\,$\langle r^2 \rangle^{3/2}$)\,meV\,=\,200.767$\pm$0.09\,meV.}    
The complete dependence on the deuteron radius is uncertain since
contributions from iteration of the potential are not included. Also,
some other contributions are not included, as indicated in table 
\ref{tab:Total-D} 

\subsubsection*{ Acknowledgments }
\vspace{-0.2cm}
The author wishes to thank M. Eides, E.-O. Le~Bigot and F. Kottmann for
extensive email correspondence regarding this work.

\small
\def\refname{{\normalsize References}}

\newpage

\subsubsection*{Appendix 1: Details of the Relativistic Recoil Calculation }
\vspace{-0.2cm}
As mentioned above, the energy levels of muonic atoms are given, to
leading order in $1/m_N$ by
\begin{equation*}
  E~=~E_r - \frac{B_0^2}{2  m_N} + \frac{1}{2 m_N} \langle h(r) +  
  2 B_0 P_1(r) \rangle
\end{equation*}
where $E_r$ is the energy level calculated using the reduced mass and 
$B_0$ is the unperturbed binding energy.  Also
\begin{equation*}
   h(r)\, = \, - P_1(r)(P_1(r) + \frac{1}{r} Q_2(r))  
            - \frac{1}{3 r} Q_2(r) [P_1(r) + Q_4(r)/r^3]
\end{equation*}
where  $P_1,\,Q_2,$\,and $Q_4$ are defined in Eq.(\ref{eq:rmp3}). 

Keeping only the Coulomb and Uehling potentials, one finds 
\begin{align*}
 P_1(r) &\,=\,- \alpha Z \frac{2 \alpha}{3\pi} (2 m_e) \chi_0(2 m_e r) \\
 Q_2(r) & \,=\, \alpha Z \left(1 + \frac{2 \alpha}{3\pi}[\chi_1(2 m_e r)
 +  (2 m_e r) \chi_0(2 m_e r)] \right)  \\
 Q_4(r) &\,=\,  \alpha Z \frac{2 \alpha}{3\pi} 
\int_1^{\infty} dz \frac{(z^2-1)^{1/2}}{z^2}  
\left(1+\frac{1}{2 z^2}\right) \\   & \cdot \left( \frac{2}{\pi} \right) 
 \int_0^{\infty} \frac{1}{q^2 + 4 m_e^2 z^2} \frac{(6qr-(qr)^3)\cos(qr)
   + (3(qr)^2-6)\sin(qr)}{q}  \,dq 
\end{align*}
\noindent where $\chi_n(x)$ is defined in \cite{RMP}.  

Since vacuum polarization is assumed to be a relatively small correction
to the Coulomb potential, it will be sufficient to approximate
$Q_2(r)$ by $\alpha Z/r $.   
After some algebra, one can reduce the expectation values to single
integrals: 
\begin{equation*} \begin{split}
\langle  P_1(r) \rangle \,=\, &  2 m_e \alpha Z \frac{2 \alpha}{3 \pi}
 \int_1^{\infty} \frac{(z^2-1)^{1/2}}{z}\cdot \left(1+\frac{1}{2
 z^2}\right) \cdot \\ & ~~~~~~~~~~~~~~~~~\left(\frac{(az)^2-az+1}{(1+az)^5}  
 \delta_{\ell 0} +\frac{1}{(1+az)^5}\delta_{\ell 1} \right)
 \,dz 
\end{split} \end{equation*}
When multiplied by $-2 B_0/(m_{\mu}+m_N)$ this results in a shift of
-0.00015\,meV for the 2s-state and of -0.00001\,meV for the 2p-state.
 For muonic deuterium, the corresponding numbers are
-0.000176\,meV and -0.000030\,meV, respectively.

\begin{equation*} \begin{split}
\langle \frac{\alpha Z}{r} P_1(r) \rangle \,=\,
 &   - (\alpha Z)^3  m_r m_e \frac{2 \alpha}{3 \pi}
 \int_1^{\infty} \frac{(z^2-1)^{1/2}}{z}\cdot \left(1+\frac{1}{2
 z^2}\right) \cdot \\ & ~~~~~~~~~~~~~~~~~\left(\frac{2(az)^2+1}{2 (1+az)^4}  
 \delta_{\ell 0} +\frac{1}{2(1+az)^4}\delta_{\ell 1} \right)
 \,dz 
\end{split} \end{equation*}
When multiplied by $1/(m_{\mu}+m_N)$ this results in a shift of
0.00489\,meV for the 2s-state and of 0.00017\,meV for the 2p-state of
muonic hydrogen.  For muonic deuterium, the corresponding numbers are
0.005543\,meV and 0.000206\,meV, respectively.

These expectation values also appear when vacuum polarization is
included in the Breit equation.

Finally,
\begin{equation*} \begin{split}
\langle \frac{\alpha Z}{3 r} Q_4(r) \rangle \,=\,
 &   - \frac{(\alpha Z)^4  m_r^2}{6} \frac{2 \alpha}{3 \pi}
 \int_1^{\infty} \frac{(z^2-1)^{1/2}}{z^2}\cdot \left(1+\frac{1}{2
 z^2}\right) \cdot \\ & ~~~~~~~~~~~~~~~\Biggl[ \Bigl[-\frac{6}{az}
 \bigl(\frac{2 +az}{1+az}-\frac{2}{az} \ln(1+az) \bigr) +
 \frac{3(az)^2+2az-1}{(1+az)^3} + \\ &~~~~~~~~~~~~~~ \frac{3+az}{4(1+az)^4} \Bigr]  
 \delta_{\ell 0} + \frac{1-3az-2(az)^2}{4(1+az)^4} \delta_{\ell 1} \Biggr]
 \,dz 
\end{split} \end{equation*}
When multiplied by $1/(m_{\mu}+m_N)$ this results in a shift of
0.002475\,meV for the 2s-state and of 0.000238\,meV for the 2p-state.
 For muonic deuterium, the corresponding numbers are
0.002753\,meV and 0.000281\,meV, respectively.

Combining these expectation values according to equations \ref{eq:rmp1}
and \ref{eq:rmp2}, one finds a contribution to the 2p-2s transition of 
-0.00419\,meV (hydrogen) and -0.00479\,meV (deuterium).  
To obtain the full relativistic and recoil corrections,
one must add the difference between the expectation values of the
Uehling potential calculated with relativistic and nonrelativistic wave
functions, 
giving a total correction of 0.0166\,meV for muonic hydrogen.  This is
in quite good agreement with the correction of .0169\,meV calculated by 
Veitia and Pachucki \cite{Pachuki3}.  The treatment presented here has the
advantage of avoiding second order perturbation theory.
For deuterium, one obtains a total correction of 0.0179\,meV.  \\


\subsubsection*{Appendix 2: Details of Corrections to the Hyperfine
  Structure of the 2s-state of  Muonic Hydrogen and Deuterium }
\vspace{-0.2cm}
The expectation value of $V_{s_1 s_2}$ in an ns state with $j=1/2$ is 
\begin{displaymath}
\Delta E_{ns}=\,=\,\frac{2 \mu_2 \alpha (\alpha Z m_r)^3}
{3 n^3 m_{\mu} m_2 s_2} \cdot (1 + a_{\mu})  [F(F+1)-s_2(s_2+1)-3/4]
\end{displaymath}
As was shown in \cite{hyperfine,eides}, the energy shift of the 2s-state 
has to be multiplied  by: 
\begin{equation*}
 1+\varepsilon_{VP} +
 \varepsilon_{vertex} + \varepsilon_{Breit} + \varepsilon_{FS,rec} 
\end{equation*}

\noindent  Here (\cite{Brodsky})
\begin{displaymath}
\varepsilon_{vertex} \,=\, \frac{2 \alpha (\alpha Z)}{3} 
\left(\ln(2)-\frac{13}{4}\right) \,=\, -1.36 \cdot 10^{-4}
\end{displaymath}
\noindent  and (\cite{eides}, Eq.\,(277))
\begin{displaymath}
\varepsilon_{Breit} \,=\, \frac{17 (\alpha Z)^2}{8} 
  \,=\, 1.13 \cdot 10^{-4}
\end{displaymath}
The vacuum polarization correction has two contributions.  One of these
is a result of a modification of the magnetic interaction between the
muon and the nucleus and is given by (see \cite{BorieHe3})
\begin{equation} \begin{split}
\varepsilon_{VP1} \,=\,& \frac{4 \alpha}{3 \pi^2}  
 \int_0^{\infty} r^2 \,dr \left(\frac{R_{ns}(r)}{R_{ns}(0)}\right)^2
 \int_0^{\infty} q^4 j_0(qr) G_M(q) \,dq   \\
 & \int_1^{\infty} \frac{(z^2-1)^{1/2}}{z^2}\cdot
\left(1+\frac{1}{2 z^2}\right) \cdot \frac{dz}{4 m_e^2 [z^2 + (q/2 m_e)^2]} 
\end{split}
\end{equation}
One can do two of the integrals analytically and obtains for the
2s-state (with
\mbox{$a=2m_e/(\alpha Z m_r)$} and  \mbox{$\sinh(\phi) = q/(2 m_e) = K/a$})  
\begin{equation}
\varepsilon_{VP1} \,=\, \frac{4 \alpha}{3 \pi^2}  
 \int_0^{\infty} \frac{K^2}{(1+K^2)^2} F(\phi) G_M(\alpha Z m_r K) \,dK 
\left[2 -\frac{7}{(1+K^2)}+\frac{6}{(1+K^2)^2}\right]
\label{eq:epsvp1}
\end{equation}
where $F(\phi)$ is known from the Fourier transform of the Uehling
potential (given as $U_2(q)$ in Ref.\,\cite{RMP}) and is given by 
\begin{equation*}%
 F(\phi) \,=\, \frac{1}{3} + (\coth^2(\phi)-3) \cdot 
  [1 + \phi \cdot \coth(\phi)]\,=\, \frac{3 \pi}{\alpha} U_2(q)
\end{equation*}%
\noindent with $\sinh(\phi)=q/2 m_e$.

The other contribution, as discussed by \cite{Brodsky,sternheim} arises
from the 
fact that the lower energy hyperfine state, being more tightly bound,
has a higher probability of being in a region where vacuum polarization
is large.  This results in an additional energy shift of  
\begin{displaymath}
 2 \int V_{Uehl}(r) \psi_{2s}(r) \delta_M \psi_{2s}(r) d^3r
\end{displaymath}
Following Ref.\,\cite{Brodsky} with $y=(\alpha Z m_r/2) \cdot r$, one has 
\begin{displaymath}
   \delta_M \psi_{2s}(r) \,=\, 2 m_{\mu} \Delta \nu_F \psi_{2s}(0) 
\left(\frac{2}{\alpha Z m_r}\right)^2 \exp(-y) 
\left[(1-y)(\ln(2y)+\gamma)+\frac{13y-3-2y^2}{4}-\frac{1}{4y}\right]
\end{displaymath}
\noindent ($\gamma$ is Euler's constant), and
\begin{displaymath}
  \psi_{2s}(r)\, =\, \psi_{2s}(0) (1-y) \exp(-y)
\end{displaymath}
 One finds after a lengthy integration
\begin{multline}
\varepsilon_{VP2} \,=\, \frac{16 \alpha}{3 \pi^2} 
 \int_0^{\infty} \frac{dK}{1+K^2}  G_E(\alpha Z m_r K)  F(\phi) \\
\biggl\{
 \frac{1}{2}-\frac{17}{(1+K^2)^2}+\frac{41}{(1+K^2)^3}-\frac{24}{(1+K^2)^4} \\
+\frac{\ln(1+K^2)}{1+K^2} \left[2-\frac{7}{(1+K^2)}+\frac{6}{(1+K^2)^2}\right]
 \\  + \frac{\tan ^{-1}(K)}{K}
 \biggl[1-\frac{19}{2(1+K^2)}+\frac{20}{(1+K^2)^2}-\frac{12}{(1+K^2)^3}\biggr] 
 \biggr\}
\label{eq:epsvp2}
\end{multline}
Sternheim\cite{sternheim} denotes the two contributions by $\delta_M$
and  $\delta_E$, respectively.
An alternative exression, obtained by assuming a point nucleus, using
Eq.(131)  from \cite{RMP} for the Uehling potential, and doing the
integrations in  a different order, is  
\begin{equation} \begin{split}
\varepsilon_{VP2} \,=\,& \frac{16 \alpha}{3 \pi} 
 \int_1^{\infty} \frac{(z^2-1)^{1/2}}{z^2}\cdot
  \left(1+\frac{1}{2 z^2}\right) \cdot \frac{1}{(1+az)^2}   \\
& \cdot \biggl[\frac{az}{2}-\frac{1}{1+az}+\frac{23}{8(1+az)^2}- \frac{3}{2(1+az)^3}   \\ 
&  +\ln(1+az)\cdot\left(1-\frac{2}{1+az}+\frac{3}{2(1+az)^2}\right)\biggr] 
\,dz 
\end{split}
\end{equation}
\noindent  with $a \,=\,2 m_e /(\alpha Z m_{red})$.
Both methods give the same result. \\
In the case of ordinary hydrogen, each of these contributes 
$3 \alpha^2/8 = 1.997 \cdot 10^{-5}$.  The accuracy of the numerical
integration was checked by reproducing these results.
One can thus expect that muonic vacuum polarization will contribute 
$3 \alpha^2/4 \simeq 4 \cdot 10^{-5}$, as in the case of normal
hydrogen.  This amounts to an energy shift of 0.0009\,meV in muonic
hydrogen and 0.0002\,meV in muonic deuterium.  
Contributions due to the weak interaction or hadronic vacuum
polarization should be even smaller.  
For muonic hydrogen, one obtains 
 $\varepsilon_{VP1}$=0.00211 and  $\varepsilon_{VP2}$=0.00325 for a
 point nucleus. Including the effect of the proton
 size (with $G_E(q)=G_M(q)$ as a dipole form factor) reduces these
 numbers to  0.00206 and 0.00321, respectively.    For the
 case of muonic deuterium, the corresponding numbers are 
 $\varepsilon_{VP1}$=0.00218 (0.00207) and  $\varepsilon_{VP2}$=0.00337
 (0.00326), respectively.   
The contribution to the hyperfine splitting of the 2s-state of hydrogen 
is then 0.0470\,meV+0.0733\,meV=0.1203\,meV (0.1212\,meV if muonic vacuum
 polarization is included).  The combined Breit and vertex
 corrections reduce this value to 0.1207\,meV.  (0.1226\,meV if the
 proton form factors are not taken into account). 

The contribution to the hyperfine structure from the two loop diagrams     
\cite{kaellen} can be calculated by replacing 
$U_2(\alpha Z m_r K) = (\alpha / 3\pi)  F(\phi) $ by 
$U_4(\alpha Z m_r K)$ (as given in \cite{RMP,Borie75}) in equations 
\ref{eq:epsvp1} and \ref{eq:epsvp2}.  The resulting contributions are  
$1.64 \cdot 10^{-5}$ and $2.46 \cdot 10^{-5}$  (for
deuterium $1.69 \cdot 10^{-5}$ and $2.54 \cdot 10^{-5}$), respectively, 
giving a total shift of 0.0009\,meV in muonic hydrogen 
and 0.0002\,meV in muonic deuterium.  

\end{document}